\documentclass[prl,aps,twocolumn,superscriptaddress,showpacs,preprintnumbers,
               amsmath,amssymb,floatfix]{revtex4}

\usepackage{graphicx} 
\usepackage{dcolumn}  
\usepackage{bm}       

\begin{document}

\title{Search for Axions with the CDMS Experiment}

\affiliation{Department of Physics, California Institute of Technology, Pasadena, CA 91125, USA}
\affiliation{Department of Physics, Case Western Reserve University, Cleveland, OH  44106, USA}
\affiliation{Fermi National Accelerator Laboratory, Batavia, IL 60510, USA}
\affiliation{Lawrence Berkeley National Laboratory, Berkeley, CA 94720, USA}
\affiliation{Department of Physics, Massachusetts Institute of Technology, Cambridge, MA 02139, USA}
\affiliation{Department of Physics, Queen's University, Kingston, ON, Canada, K7L 3N6}
\affiliation{Department of Physics, Saint Olaf College, Northfield, MN  55057}
\affiliation{Department of Physics, Santa Clara University, Santa Clara, CA 95053, USA}
\affiliation{Department of Physics, Stanford University, Stanford, CA 94305, USA}
\affiliation{Department of Physics, Syracuse University, Syracuse, NY 13244}
\affiliation{Department of Physics, Texas A\&M University, College Station, TX 93106, USA} 
\affiliation{Department of Physics, University of California, Berkeley, CA 94720, USA}
\affiliation{Department of Physics, University of California, Santa Barbara, CA 93106, USA}
\affiliation{Departments of Phys. \& Elec. Engr., University of Colorado Denver, Denver, CO 80217, USA}
\affiliation{Department of Physics, University of Florida, Gainesville, FL 32611, USA}
\affiliation{School of Physics \& Astronomy, University of Minnesota, Minneapolis, MN 55455, USA}
\affiliation{Physics Institute, University of Z\"urich, Z\"urich, Switzerland}

\author{Z.~Ahmed} \affiliation{Department of Physics, California Institute of Technology, Pasadena, CA 91125, USA}
\author{D.S.~Akerib} \affiliation{Department of Physics, Case Western Reserve University, Cleveland, OH  44106, USA} 
\author{S.~Arrenberg} \affiliation{Physics Institute, University of Z\"urich, Z\"urich, Switzerland} 
\author{C.N.~Bailey} \affiliation{Department of Physics, Case Western Reserve University, Cleveland, OH  44106, USA} 
\author{D.~Balakishiyeva}\affiliation{Department of Physics, University of Florida, Gainesville, FL 32611, USA}
\author{L.~Baudis} \affiliation{Physics Institute, University of Z\"urich, Z\"urich, Switzerland} 
\author{D.A.~Bauer} \affiliation{Fermi National Accelerator Laboratory, Batavia, IL 60510, USA} 
\author{J.~Beaty} \affiliation{School of Physics \& Astronomy, University of Minnesota, Minneapolis, MN 55455, USA} 
\author{P.L.~Brink} \affiliation{Department of Physics, Stanford University, Stanford, CA 94305, USA} 
\author{T.~Bruch} \affiliation{Physics Institute, University of Z\"urich, Z\"urich, Switzerland} 
\author{R.~Bunker} \affiliation{Department of Physics, University of California, Santa Barbara, CA 93106, USA} 
\author{B.~Cabrera} \affiliation{Department of Physics, Stanford University, Stanford, CA 94305, USA} 
\author{D.O.~Caldwell} \affiliation{Department of Physics, University of California, Santa Barbara, CA 93106, USA} 
\author{J.~Cooley} \affiliation{Department of Physics, Stanford University, Stanford, CA 94305, USA} 
\author{P.~Cushman} \affiliation{School of Physics \& Astronomy, University of Minnesota, Minneapolis, MN 55455, USA} 
\author{F.~DeJongh} \affiliation{Fermi National Accelerator Laboratory, Batavia, IL 60510, USA} 
\author{M.R.~Dragowsky} \affiliation{Department of Physics, Case Western Reserve University, Cleveland, OH  44106, USA} 
\author{L.~Duong} \affiliation{School of Physics \& Astronomy, University of Minnesota, Minneapolis, MN 55455, USA} 
\author{E.~Figueroa-Feliciano}\affiliation{Department of Physics, Massachusetts Institute of Technology, Cambridge, MA 02139, USA}
\author{J.~Filippini} \affiliation{Department of Physics, University of California, Berkeley, CA 94720, USA}\affiliation{Department of Physics, California Institute of Technology, Pasadena, CA 91125, USA} 
\author{M.~Fritts} \affiliation{School of Physics \& Astronomy, University of Minnesota, Minneapolis, MN 55455, USA} 
\author{S.R.~Golwala} \affiliation{Department of Physics, California Institute of Technology, Pasadena, CA 91125, USA} 
\author{D.R.~Grant} \affiliation{Department of Physics, Case Western Reserve University, Cleveland, OH  44106, USA} 
\author{J.~Hall} \affiliation{Fermi National Accelerator Laboratory, Batavia, IL 60510, USA} 
\author{R.~Hennings-Yeomans} \affiliation{Department of Physics, Case Western Reserve University, Cleveland, OH  44106, USA} 
\author{S.~Hertel}\affiliation{Department of Physics, Massachusetts Institute of Technology, Cambridge, MA 02139, USA}
\author{D.~Holmgren} \affiliation{Fermi National Accelerator Laboratory, Batavia, IL 60510, USA} 
\author{L.~Hsu} \affiliation{Fermi National Accelerator Laboratory, Batavia, IL 60510, USA} 
\author{M.E.~Huber} \affiliation{Departments of Phys. \& Elec. Engr., University of Colorado Denver, Denver, CO 80217, USA} 
\author{S.W.~Leman} \affiliation{Department of Physics, Massachusetts Institute of Technology, Cambridge, MA 02139, USA} 
\author{R.~Mahapatra}\affiliation{Department of Physics, Texas A\&M University, College Station, TX 93106, USA} 
\author{V.~Mandic} \affiliation{School of Physics \& Astronomy, University of Minnesota, Minneapolis, MN 55455, USA} 
\author{D.~Moore} \affiliation{Department of Physics, California Institute of Technology, Pasadena, CA 91125, USA}
\author{K.A.~McCarthy}\affiliation{Department of Physics, Massachusetts Institute of Technology, Cambridge, MA 02139, USA}
\author{N.~Mirabolfathi} \affiliation{Department of Physics, University of California, Berkeley, CA 94720, USA} 
\author{H.~Nelson} \affiliation{Department of Physics, University of California, Santa Barbara, CA 93106, USA} 
\author{R.W.~Ogburn} \affiliation{Department of Physics, Stanford University, Stanford, CA 94305, USA}\affiliation{Department of Physics, California Institute of Technology, Pasadena, CA 91125, USA} 
\author{M.~Pyle} \affiliation{Department of Physics, Stanford University, Stanford, CA 94305, USA}
\author{X.~Qiu} \affiliation{School of Physics \& Astronomy, University of Minnesota, Minneapolis, MN 55455, USA} 
\author{E.~Ramberg} \affiliation{Fermi National Accelerator Laboratory, Batavia, IL 60510, USA} 
\author{W.~Rau} \affiliation{Department of Physics, Queen's University, Kingston, ON, Canada, K7L 3N6}
\author{A.~Reisetter} \affiliation{Department of Physics, Saint Olaf College, Northfield, MN  55057}\affiliation{School of Physics \& Astronomy, University of Minnesota, Minneapolis, MN 55455, USA}
\author{T.~Saab}\affiliation{Department of Physics, University of Florida, Gainesville, FL 32611, USA}
\author{B.~Sadoulet} \affiliation{Lawrence Berkeley National Laboratory, Berkeley, CA 94720, USA}\affiliation{Department of Physics, University of California, Berkeley, CA 94720, USA}
\author{J.~Sander} \affiliation{Department of Physics, University of California, Santa Barbara, CA 93106, USA} 
\author{R.W.~Schnee} \affiliation{Department of Physics, Syracuse University, Syracuse, NY 13244} 
\author{D.N.~Seitz} \affiliation{Department of Physics, University of California, Berkeley, CA 94720, USA} 
\author{B.~Serfass} \affiliation{Department of Physics, University of California, Berkeley, CA 94720, USA} 
\author{K.M.~Sundqvist} \affiliation{Department of Physics, University of California, Berkeley, CA 94720, USA} 
\author{M.~Tarka} \affiliation{Physics Institute, University of Z\"urich, Z\"urich, Switzerland} 
\author{G.~Wang} \affiliation{Department of Physics, California Institute of Technology, Pasadena, CA 91125, USA}
\author{S.~Yellin} \affiliation{Department of Physics, Stanford University, Stanford, CA 94305, USA} \affiliation{Department of Physics, University of California, Santa Barbara, CA 93106, USA}
\author{J.~Yoo} \affiliation{Fermi National Accelerator Laboratory, Batavia, IL 60510, USA} 
\author{B.A.~Young} \affiliation{Department of Physics, Santa Clara University, Santa Clara, CA 95053, USA} 

\collaboration{CDMS Collaboration}
\noaffiliation

\begin{abstract}
 We report on the first axion search results from the Cryogenic Dark Matter Search (CDMS) experiment at the Soudan Underground Laboratory. An energy threshold of 2\,keV for electron-recoil events allows a search for possible solar axion conversion into photons or local Galactic axion conversion into electrons in the germanium crystal detectors. The solar axion search sets an upper limit on the Primakov coupling $g_{a\gamma\gamma}$ of 2.4$\times10^{-9}$\,GeV$^{-1}$ at the 95\% confidence level for an axion mass less than 0.1\,keV/c$^2$. This limit benefits from the first precise measurement of the absolute crystal plane orientations in this type of experiment. The Galactic axion search analysis sets a world-leading experimental upper limit on the axio-electric coupling $g_{a\bar{e}e}$ of 1.4$\times10^{-12}$ at the 90\% confidence level for an axion mass of 2.5\,keV/c$^2$. This analysis excludes an interpretation of the DAMA annual modulation result in terms of Galactic axion interactions for axion masses above 1.4\,keV/c$^2$. 
\end{abstract}
\preprint{FERMILAB-PUB-09-053-E}
\maketitle

The axion has been postulated to solve the strong CP problem in quantum chromodynamics. The breaking of Peccei-Quinn U(1) symmetry leaves a pseudo-Goldstone boson field~\cite{peccei1977}, interpreted as  the axion. Although the original Peccei-Quinn axion model has been ruled out~\cite{weinberg1978, wilczek1978}, ``invisible'' axion models allow a wide range of axion masses and axion-matter couplings~\cite{jekim1987}. Astrophysical observations are currently the best strategy to search for these invisible axions~\cite{sikivie1983}. The interior of stars is expected to be a powerful source of axions due to the high abundance of photons and strong electromagnetic fields, which may convert photons into axions. The non-thermal axion production mechanism in the early universe provides a cold dark matter candidate.

Here we report on the first axion search results from the Cryogenic Dark Matter Search (CDMS) experiment. The CDMS collaboration operates a total of 19\ Ge ($\sim$250\,g each) and
11\ Si ($\sim$100\,g each) crystal detectors at $\sim$\,40\,mK in the Soudan Underground Laboratory. The detectors are designed to read out both ionization and phonon signals from recoil events~\cite{zips}. The ratio of ionization to phonon energy, the ionization yield, enables discrimination of nuclear from electron recoils. The details of the detector structure and operation can be found in reference~\cite{prd118}. We report on data from two run periods between October 2006 and July 2007, also used for the WIMP-search analysis described in~\cite{cdms2008}. This analysis modifies our event selection criteria to focus on electron-recoil events, lowering the energy threshold to 2\,keV for such events and giving a net Ge exposure of 443.2\,kg-days before cuts.

The flux of solar axions at the Earth can be estimated assuming the standard solar model~\cite{bp2004} and a coupling to the $\cal O$(keV) black body photons in the core region of the Sun. For axion masses $\ll$ 1 keV/c$^2$, photon-axion conversion creates a flux of $\cal O$(keV) axions at the Earth. The solar axion flux at the Earth is given by~\cite{cast2007, cast2008} :
\begin{equation}
  \frac{d\Phi_a}{dE_a} = \frac{6.02\cdot10^{14}}{ \text{ cm}^{2} \text{ s} \text{ keV}}\left(\frac{g_{a\gamma\gamma}\cdot10^{8}}{\text{GeV}^{-1}}\right)^2 E_a^{2.481}e^{-E_a/1.205},
\end{equation}
\noindent
where $E_a$ is the energy of the axion in keV and $g_{a\gamma\gamma}$ is the axion-photon coupling constant.

The axion-photon coupling to the nuclear Coulomb field in the detectors converts
axions back into photons of the same energy (Primakov effect). Coherent Bragg diffraction produces
 a strong correlation between incident beam direction and conversion probability, providing a unique signature of solar axions. The expected event rate can be computed as a function of energy and the orientation of the crystal relative to the location of the Sun \cite{pascos1994}. 
It is given as a function of observed photon energy $E$ for a given axion momentum transfer $\vec{q}$ and scattering angle $\theta$~\cite{cebrian1999}, by:
\begin{eqnarray} \label{fcn:rate}
\begin{array}{l}
    \displaystyle \mathcal{R}(E) = \displaystyle 2c \int \frac{d^3q}{q^2}\frac{d\Phi_a}{dE_a} \left[\frac{g^2_{a\gamma\gamma}}{16\pi^2} |F(\vec{q})|^2 \sin^2(2\theta)\right]{\cal{W}},
\end{array}
\end{eqnarray}
\noindent
where ${\cal{W}}$ is a detector energy resolution function. The Fourier transform of the electric field in a crystal is given as $F(\vec{q}) = k^2 \int d^3x \phi(\vec{x}) e^{i\vec{q}\cdot\vec{x}}$, which depends on $\phi(\vec{x}) = \sum_i \frac{Z e}{4\pi|\vec{x}-\vec{x}_i|}e^{-\frac{|\vec{x}-\vec{x}_i|}{r}} = \sum_G n_G e^{i\vec{G}\cdot\vec{x}}$, where $k$ is the photon momentum, $e$ is the elementary charge, $\vec{x}_i$ is the position of a germanium atom in the lattice, $r$ is the screening length of the atomic electric field, $Z$=32 for germanium, and $\vec{G}$ is a reciprocal lattice vector. The structure coefficients $n_G$ (defined in \cite{cebrian1999}) account for the face-centered-cubic structure of Ge. The Bragg condition ($\vec{q}=\vec{G}$) can be expressed in terms of the axion energy as $E_a = \hbar c |\vec{G}|^2 / (2 \hat{u}\cdot\vec{G})$, where $\hat{u}$ is a unit vector directed towards the Sun.

The expected event rate is calculated based on an accurate measurement of the orientation of each detector with respect to the position of the Sun. We took the specific geometry of the experiment, the live-time during data taking and the seasonal modulation of the solar axion flux due to the changing distance between the Sun and the Earth into account.
The geodesic location of the Soudan Underground Laboratory is latitude 47.815$^{\circ}$N, longitude 92.237$^{\circ}$W and altitude 210\,m below sea level. The geodesic north of the CDMS experimental cavern was measured in 1999 by the Fermilab Alignment Group~\cite{alignment1999}. A line connecting two survey points along the central axis of the cavern was found to be 0.165$^{\circ}$E from true north. By extension, the main horizontal axis of the CDMS cryostat was found to be 0.860 $\pm$ 0.018$^{\circ}$E from true north. 

\begin{figure}[t!]
    \includegraphics[width=3.2in]{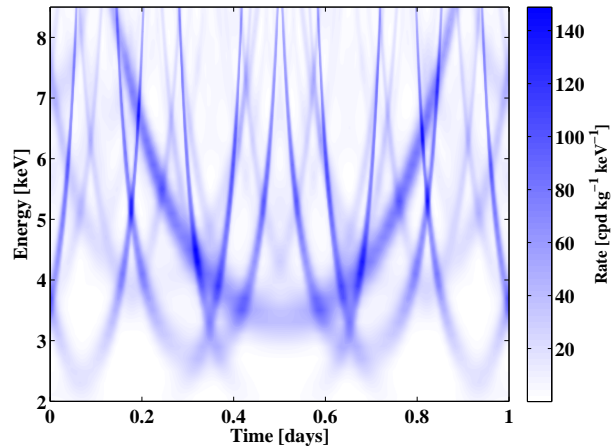} 
    \caption{\small Time and energy dependence of the expected solar axion conversion rate in a Ge detector for $g_{a\gamma\gamma}=10^{-8} \text {GeV}^{-1}$.}
\label{fig:solaxrate}
\end{figure}

Within the cryostat the 30 CDMS detectors are mounted in five towers of six detectors each. The vertical axis of each tower is aligned with the [001] axis of the detectors. The (110) axis that defines the major flat on each substrate is rotated with respect to its neighbors above and below, such that the detectors form a helix within each tower. The uncertainty in the absolute azimuth orientation of the crystal planes is dominated by an estimated 3$^{\circ}$ uncertainty in the exact angular position of the tower axes with respect to the central axis of the cryostat. The uncertainty of the zenith angle measurement was estimated to be less than 1$^{\circ}$. 
In Fig.~\ref{fig:solaxrate} we present the predicted event rate in a germanium detector for an assumed 
coupling of $g_{a\gamma\gamma}=10^{-8}\text{ GeV}^{-1}$. 

\begin{figure}[t!]
\includegraphics[width=3.1in]{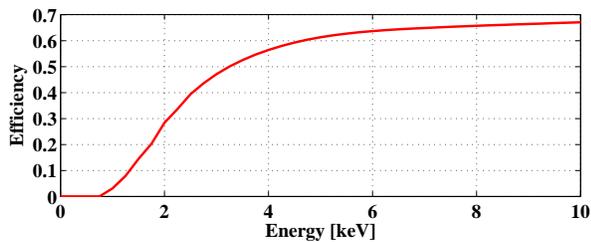}
\caption{\small Detection efficiency as a function of energy.}
\label{fig:deteff}
\end{figure}

In order to sample pure axion interaction candidate events, we require that an event ($>$\,3$\sigma$ above mean noise) was recorded in one and only one detector; i.e., a single scatter. To make sure the selected events are not due to residual cosmic ray interactions, they are required not to be coincident in time with activity in the veto shield surrounding the apparatus. Candidate events are selected within the $\pm$2$\sigma$ region of the electron-recoil distribution in ionization yield. Data sets taken within 3 days after neutron calibrations are not considered in order to avoid high gamma rates due to activation. The detection efficiency is dominated by rejection of events with an ionization signal in a detector annular guard electrode, covering 15\% of the detector volume. The acceptance of our analysis cuts for single-scatter electron recoils was measured as a function of energy (Fig.~\ref{fig:deteff}). The average event rate of electron recoil singles below 100\,keV in all detectors is stable in time at 16\%.

\begin{figure}[t!]
  \includegraphics[width=3.2in]{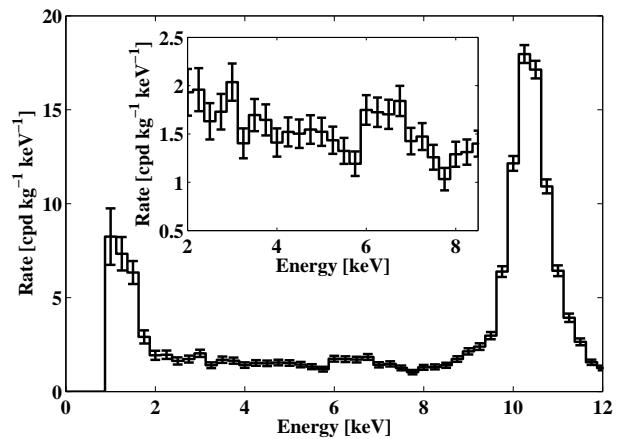}
  \caption{\small Co-added, efficiency corrected low energy spectrum of the Ge detectors considered in this analysis. The inset shows an enlargement of the spectrum in the analysis window, taken to be 2 to 8.5\,keV.}
  \label{fig:enespc}
\end{figure}

For the germanium detectors considered in this analysis, the summed background rate after correcting for detection efficiency is $\sim$1.5 cpd (counts per day) kg$^{-1}$ keV$^{-1}$ (Fig.~\ref{fig:enespc}). The prominent 10.36\,keV line is caused by X-rays and Auger-electrons from the electron-capture decay of $^{71}$Ge, produced by neutron capture on $^{70}$Ge during $^{252}$Cf calibration of the detectors. The excess in event rate around 6.5\,keV (inset) is likely caused by remnant $^{55}$Fe decays from cosmogenic activation. The de-excitation of $^{55}$Mn following the electron-capture decay of $^{55}$Fe yields a total of 6.54\,keV of electron-recoil events. We interpolate the energy resolution of the 10.36\,keV line (typically better than 3\%) to the noise level to obtain the energy-dependent resolution of each detector. The analysis window, defined from 2 to 8.5\,keV, is determined by the expected axion flux, background rate, and detection efficiency.

\begin{figure}[t!]
   \includegraphics[width=3.2in]{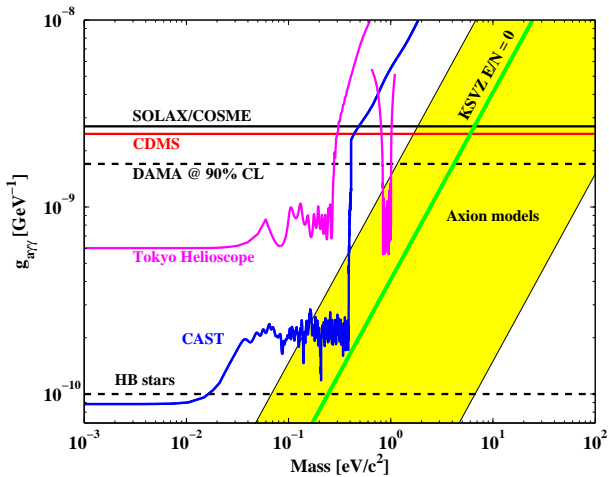}
    \caption{\small Comparison of the 95\% C.L. upper limit on $g_{a\gamma\gamma}$ achieved in this analysis (red/solid) with other crystal search experiments (SOLAX/COSME~\cite{solax1998,cosme2001} (black/solid) and DAMA (upper black/dashed)~\cite{dama2001solaxion}) and helioscopes (Tokyo heslioscope (magenta/solid)~\cite{tokyo2008} and CAST (blue/solid)~\cite{cast2008}). The constraint from Horizontal Branch stars (lower black/dashed) is also shown~\cite{raffelt2008}. }
    \label{fig:gagg_limit}
 \end{figure}

We carried out a profile likelihood analysis to determine the best fit value of $g_{a\gamma\gamma}$.
We express the event rate per unit measured energy ($E$), per unit time ($t$), and per detector ($d$) of a solar axion signal with background as 
\begin{eqnarray}
  R(E,t,d) = \varepsilon(E,d) \left[\lambda \mathcal{R}(E,t,d) + B(E,d) \right],
\label{eqn:rate}
\end{eqnarray}
\noindent
where $\varepsilon(E,d)$ is the detection efficiency, $\mathcal{R}(E,t,d)$ is the expected event rate for a coupling constant $g_{a\gamma\gamma} = 10^{-8}\text{ GeV}^{-1}$, and $\lambda = (g_{a\gamma\gamma} \cdot 10^{8}\text{ GeV})^4$ is the scale factor for the actual value of $g_{a\gamma\gamma}$. $B(E,d)$ is the background described by
\begin{eqnarray}
 \nonumber
B(E,d)&\equiv& C(d)+D(d)E+H(d)/E \\
     &+& \frac{\eta_{6.54}}{\sqrt{2 \pi} \sigma_{6.54}} e^{\left(-\frac{(E-6.54\text{keV})^2}{2 \sigma_{6.54}^2}\right)},
\end{eqnarray}
\noindent
where $C(d)$, $D(d)$, and $H(d)$ are free parameters. The Gaussian term describes a contribution from $^{55}$Fe decays at an energy of 6.54\,keV and unknown total rate $\eta_{6.54}$. The fitting is done by maximizing the unbinned log likelihood function with respect to $\lambda$ and the background parameters, for individual events {\it i}:
\begin{equation}  \label{eqn:loglike}
  log(\mathcal{L}) = - R_T +\sideset{}{_{i,j}} \sum log(R(E_i,t_i,d_j)),
\end{equation}
\noindent
where $R_T$ is the total sum of the event rate ($R$) over energy, time, and detectors. The scaling factor from the maximization $\lambda=(1 \pm 1.5) \times 10^{-3} $ is compatible with zero. No indication of solar axion conversion to photons is observed. Given a null observation, we set an upper limit on the coupling constant $g_{a \gamma \gamma}$, where the scaling factor $\lambda$ is obtained by integrating the profile likelihood in the physically allowed region ($\lambda > 0$). The upper limit on the axion-photon coupling, $g_{a\gamma\gamma} < 2.4 \times10^{-9}$\,GeV$^{-1}$ at a 95\% C.L. is the only laboratory bound based on the accurate measurement of all crystal orientations of the detectors. None of the previous solar axion search experiments (SOLAX/COSME/DAMA) measured their crystal orientations~\cite{solax1998,cosme2001,dama2001solaxion}, and thus their limits are penalized by picking the least sensitive direction for their solar axion bound. The result of this analysis is compared to other experimental constraints in Fig.~\ref{fig:gagg_limit}. Improvement towards the next order of sensitivity requires improvements in both detector exposure and gamma background level. A 100-kg SuperCDMS experiment, with substantially reduced gamma background level ($\sim$0.1~cpd kg$^{-1}$ keV$^{-1}$) would improve the sensitivity to $g_{a\gamma\gamma} < 10^{-9}$\,GeV$^{-1}$. 

\par
\begin{figure}[t!]
  \includegraphics[width=3.2in]{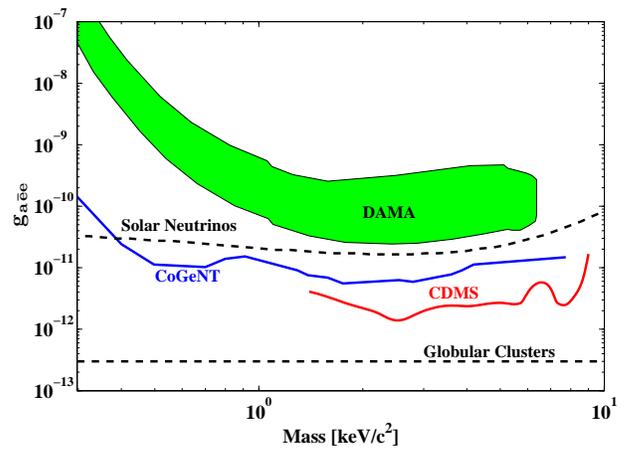}
  \caption{\small The allowed region (green/filled) from a Galactic axion interpretation of the annual modulation signature observed by the DAMA experiment~\cite{dama2006} is shown. The 90\% C.L. upper limits on the $g_{a\bar{e}e}$ coupling constant from this work (red/solid) and the CoGeNT experiment (blue/solid)~\cite{cogent2008} completely exclude the DAMA allowed region. The indirect constraints from astrophysical bounds (black/dashed) are also shown~\cite{gondolo2008}.}
  \label{fig:gaee_limit}
\end{figure}

 In addition to restricting solar axions, the CDMS measurement can be used to limit Galactic axions. The annual modulation signature observed by DAMA may be interpreted as a detection of axions distributed in the local Galactic halo~\cite{dama2006,damalibra2008}. If present, these non-relativistic axions would materialize in our detectors via an axio-electric coupling ($g_{a\bar{e}e}$). 
Assuming a local Galactic dark matter mass density of 0.3\,GeV/c$^2$/cm$^3$, the expected event rate~\cite{pospelov2008} is given by:
\begin{equation}\label{eqn:rate_gaee}
  \text{R [cpd kg$^{-1}$]} = 1.2 \times 10^{43}A^{-1} g^2_{a\bar{e}e} m_a \sigma_{p.e},
\end{equation}
\noindent
where $m_a$ is the axion mass in keV$/c^2$, $A$=73 for germanium, and $\sigma_{p.e}$ is the photoelectric cross section in cm$^2$ per atom. We analyzed the energy spectrum using the same electron-recoil data samples used in the solar axion search, as shown in Fig.~\ref{fig:enespc}. We carried out a profile likelihood calculation to search for an excess of event rate above background. The same formalism described in equations~(\ref{eqn:rate}) to~(\ref{eqn:loglike}) was used, with the term for the expected conversion rate of solar axions $\mathcal{R}(E,t,d)$ replaced by a Gaussian distribution function representing a spectral line at a given energy or axion mass.  We find no statistically significant excess of event rate above background. Lacking a direct constraint on a possible $^{55}$Fe contribution to the spectrum, we set a conservative upper limit, shown in Fig.~\ref{fig:gaee_limit}, on the total counting rate in this energy range without any attempt to subtract a possible background contribution. This result excludes significant new Galactic axion parameter space in the mass range between 1.4 and 9\,keV/c$^2$, and is inconsistent with the interpretation of the DAMA signature due to axions ~\footnote{The DAMA region should be understood with caution. The non-relativistic speed of Galactic axions distributed in the local halo causes the conversion rate to be independent of the particle's velocity, thus the annual modulation of the counting rate is highly suppressed ~\cite{pospelov2008}. Therefore, the DAMA modulation signal is too large to be interpreted by Galactic axion interactions in the detectors.}.

\par In summary, the solar axion search sets an upper limit on the Primakov coupling $g_{a\gamma\gamma}$ of 2.4$\times10^{-9}$\,GeV$^{-1}$ at the 95\% confidence level for an axion mass less than $\sim$0.1\,keV/c$^2$. This limit is the first one based on accurate measurements of crystal orientations.
The systematic error on the limit is estimated to be 7.9\%, which arises from the remaining uncertainty in the alignment of the detector towers' major axes to the central cryogenic axis. The local Galactic axion search analysis sets a world-leading experimental upper limit on the axio-electric coupling $g_{a\bar{e}e}$ of 1.4$\times10^{-12}$ at the 90\% confidence level for an axion mass of 2.5\,keV/c$^2$, and excludes the axion interpretation of the DAMA signature in the mass range of 1.4\,keV/c$^2$ to 9\,keV/c$^2$. 

\par This experiment would not have been possible without the contributions of numerous engineers and technicians; we would like to especially thank Larry Novak, Richard Schmitt and Astrid Tomada. We thank the CAST and Tokyo helioscope collaborations for providing us with their axion limits. The direction measurement of the true north in the Soudan Underground Laboratory relied on the help from the Fermilab Alignment Group. Special thanks to Virgil Bocean.  This work is supported in part by the National Science Foundation (Grant Nos.\ AST-9978911, PHY-0542066, PHY-0503729, PHY-0503629,  PHY-0503641, PHY-0504224 and PHY-0705052), by the Department of Energy (Contracts DE-AC03-76SF00098, DE-FG02-91ER40688, DE-FG02-92ER40701, DE-FG03-90ER40569, and DE-FG03-91ER40618), by the Swiss National Foundation (SNF Grant No. 20-118119), and by NSERC Canada (Grant SAPIN 341314-07).

\end{document}